# Investigation of NOx in piloted stabilized methane-air diffusion flames using Finite-rate and Infinitely-fast chemistry based combustion models

Rohit Saini[1], Swetha Prakash[1], Ashoke De[1*], Rakesh Yadav[2]

[1]Department of Aerospace Engineering, Indian Institute of Technology, Kanpur, India - 208016

[2]ANSYS Inc., 5930 Cornerstone Court West, Suite 230 CA 92121 USA

*Corresponding Author: Tel.: +91-512-2597863 Fax: +91-512-2597561

E-mail address: ashoke@iitk.ac.in

**ABSTRACT**

The present work reports on the numerical investigation of NOx in three turbulent piloted diffusion flames of different levels of extinction. The study involves two-dimensional axisymmetric modeling of combustion in these flames with fairly detailed chemistry, i.e. GRI 3.0 mechanism. The main focus of the study is to analyze the effects of the two different combustion model approaches, such as infinitely fast chemistry based unsteady flamelet and finite rate chemistry based EDC, in predicting the NOx formation in three piloted methane jet flames (Sandia D, E, and F). The EDC approach is able to predict the passive scalar quantities but shows over-prediction in the reactive scalar quantities and NO prediction, while the unsteady flamelet modeling is found to be essential in predicting the accurate formation of slow kinetic species like NOx. The inability of flamelet and EDC approach in capturing localized flame extinction is observed, which lead to an over-prediction of NOx at larger downstream locations. Further, the dominance of NOx formation pathways is investigated in all three flames.

**NOMENCLATURE**

| | |
|---|---|
| $\rho$ | Density |
| $\phi$ | representative scalar |
| $D$ | diffusion coefficient |
| $\sim$ | Favre average |
| $Z$ | mixture fraction |
| $Z_{st}$ | stoichiometric mixture fraction |
| $Z''$ | mixture fraction variance |
| $\chi$ | scalar dissipation |
| $\chi_{st}$ | scalar dissipation at stoichiometric mixture fraction |
| $I$ | unsteady flamelet marker probability |
| $erfc^{-1}$ | inverse complementary error function |
| $Y_i$ | species mass fraction |
| $\mu_t$ | turbulent viscosity |
| $\sigma_t$ | turbulent Prandtl number |
| $C_p$ | specific heat |
| $T$ | Temperature |
| $T^{ad}$ | adiabatic flame temperature |
| $V$ | fluid volume |
| $t$ | Time |
| $w$ | Reaction rate |
| $t$ | Residence time scale |

## 1. INTRODUCTION

Combustion efficiency is exceptionally reliant on the turbulence nature and their associated levels involved during the combustion phenomena such that to alter their rate of combustion in most of the industrial burners. Combustion involves a large number of elementary reactions and may result in the emission of various pollutant species like carbon monoxide (CO), Oxides of Nitrogen (NOx) and soot (highly branched Carbon particles). Performing simulations of turbulent combustion and accurately predicting the results have thus been of concern for many researchers for a few decades now. Modeling combustion to understand the underlying relationship between turbulence and chemistry to reduce emissions and to improve combustion efficiency is important since regulations on emissions from combustion processes are becoming stringent day by day through government stringent norms [1]. Understanding the chemical and physical aspects of combustion helps in optimizing the underlying processes to result in as lowest emission as possible. Flamelet modeling is one of the most commonly used approaches for modeling turbulent combustion due to its computational efficiency [2 & 3]. Steady flamelet model can simulate local chemical non-equilibrium due to aerodynamic straining of the flame by the turbulent flow field whereas, slow kinetic species like NOx that don't respond quickly to the turbulent straining have to be modeled using unsteady flamelet model for accurate prediction since their formation depends on the history of molecular mixing and reactions and its formulation is extensively studied by [4]. [1] performed laminar flamelet modeling of turbulent jet flame and predicted NOx by solving another transport equation for NO mass fraction. This is an alternate approach to separate the chemical scales, but solving separate NO transport equation is typically associated with a frozen chemistry. Whereas, the unsteady flamelet modeling can describe the chemistry effects accurately as it considers and solves the entire reaction mechanism in a coupled manner [5]. [6] investigated NOx characteristics in turbulent non-premixed jet and recirculating flames using both Conditional Moment Closure (CMC) and unsteady flamelet approach; whereas [7] studied steady and unsteady flamelet modeling of Hydrogen/air flame and observed that the steady flamelet model over predicted the NO mass fraction by over an order of magnitude while unsteady flamelet modeling provided reasonable predictions. Recently, [8] compared EDC-based and steady laminar flamelet based results with the experimental measurements of the Sandia flame D and observed significant improvement in the reactive as well as passive scalars under EDC-based approach.

Nitric oxide (NO) and nitrogen dioxide ($NO_2$) collectively known as NOx, forms mainly in the high-temperature regions of the flame where stoichiometric combustion takes place since high energy is required to break the triple bond that binds the nitrogen atoms in the stable $N_2$ molecule. Hence the kinetics of formation of species like NOx is slow and it forms in the regions of high residence times where steady diffusion flamelet model (SDFM) approaches equilibrium solution and could over-predict NOx by orders of magnitude. Scalar dissipation rate, the parameter that represents the state of non-equilibrium of the flamelet decays strongly along the jet axis due to relaxation of strain and steady flamelet cannot follow the rapid changes of the scalar dissipation rate instantaneously [4-7] and hence it slowly tends to approach the state of equilibrium as the strain relaxes. In the present study, we resort to an unsteady flamelet modeling approach in order to accurately represent the influence of transient effects in the formation of slow kinetic species and to represent the transient history of scalar dissipation of individual flames.

Scalar dissipation rate varies with time as well as space and hence resolving the spatial dependence of scalar dissipation could influence the accuracy of predictions. The number of unsteady flamelets used decides the spatial resolution of scalar dissipation and its evolution. [9 & 10] studied pollutant formation in the unsteady reactive flows in diesel engines by solving the unsteady flamelets using Eulerian Particle Flamelet model (EPFM), where each particle represented an unsteady flamelet and observed that NOx levels were excellently reproduced with 10 flamelets. [11] also used EPFM for NO predictions in turbulent diffusion flames. Further, [12] used EPFM to study NOx formation in Sandia flame D and observed that a single unsteady flamelet was sufficient to obtain reasonable predictions. They studied the effect of different methods of initialization of the probability of finding particles and initialization region extended to fuel rich regions provided better results. [4] conducted a study on Sandia flame D and Delft flame III using multiple unsteady flamelet method and observed similar trends in NO predictions as in [12]. They also found that increasing the number of flamelets increased the accuracy of prediction in the Delft flame III whereas the results hardly improved for Sandia flame D. The present study extends the analysis on Sandia flames E and F that have not been studied before on these aspects while comparing with the finite rate chemistry-based approaches.

A number of previous investigations on turbulent diffusion flames have validated that the usage of significantly reduced chemistry could result in difficulties in flame stabilization and substantial discrepancies in the predicted mass fractions of even major species [13]. The chemical mechanism used is generally a compromise between computational efficiency and the requirement of sufficient detail in the representation of chemistry [14]. Developing improved mechanisms for methane combustion including NOx chemistry has been the subject of research for many years now. [15] used a reduced mechanism derived from GRI 1.2 (ARM 2) for methane oxidation in the PDF calculations of flames D - F and a very good agreement between the predicted results and experimental data for NOx were observed. [16] used a skeletal mechanism for methane, GRI 2.11 and Miller - Bowman mechanism to perform Conditional Moment Closure (CMC) prediction on flame D and [4] used GRI 2.11 in the unsteady flamelet modeling of pollutants in Delft Flame III and Sandia flame D and observed that GRI mechanism produces very good agreement with the experimental results for NO.

In the present work, the two different turbulence-chemistry interaction approaches are used to investigate the flame structure and NO formation with different pathways and their corresponding dominance in the net production of the NO. GRI 3.0 chemistry by [17], is used in the present study which is a successor of the 2.11 mechanism with NOx formation and reburn chemistry that includes CH kinetics important for prompt NOx formation [18] and has been used in various combustion modeling studies [18-21]. The main objective of the present study is to analyze and understand the effect of the unsteady flamelet modeling and eddy dissipation concept approach in NOx predictions. A flame with high Reynolds number and lower degree of localized flame extinction can be described by quite a wide range of combustion and turbulence models. Hence, a large bank of data is available for Sandia flame D whereas, not many studies on the combustion simulations and NOx modeling are available on the flames E and F, as per authors' knowledge. Therefore, Sandia flames series (D to F) is chosen for the present study. This study could shed light on the accuracy of multiple unsteady flamelet modeling and eddy dissipation concept on the flames with higher degrees of localized flame extinction. Radiation is neglected as an assumption to investigate the effect on NO in the Sandia flames by [22]. Whereas, [23] performed radiation measurements in flames C-F and observed that only a relatively small portion (5% and less) of the total heat release was represented by radiation. For numerical simplification, the present study utilizes an adiabatic approach towards the simulations of all flames thereby neglecting radiation. Furthermore, unsteady flamelet equations are solved as a post-processing step on the converged solution from the steady diffusion flamelet modeling with a frozen flowfield.

**MATHEMATICAL FORMULATION**

**1.1 Governing Equation**

The general form of the Favre averaged governing equation for the transport of mass, momentum, and turbulence can be expressed as:

$$\frac{\partial}{\partial t}\rho\tilde{\phi} + \frac{\partial}{\partial x_j}\rho u_j \tilde{\phi} = \frac{\partial}{\partial x_j}\rho D \frac{\partial \tilde{\phi}}{\partial x_j} + \langle S_\phi \rangle \tag{1}$$

Where $\tilde{\phi}$ represents the flow field variable solved for and $\langle S_\phi \rangle$ represents the source of $\tilde{\phi}$. The scalar diffusion coefficient and mean density are represented by $D$ and $\rho$ respectively. The turbulence model provides the turbulence scale information.

**1.2 Turbulence-Chemistry Interaction**

**1.2.1 Unsteady Diffusion Flamelet Modelling approach**

The steady flamelet model is limited to modeling combustion with fast chemistry and is assumed to respond instantaneously to the aerodynamic strain. When the chemical time scales are comparable with the mixing time scales, the chemistry remains at non-equilibrium state since the species do not have sufficient time to reach chemical equilibrium. Hence steady flamelets cannot accurately predict slow chemistry and transient effects such as ignition and extinction due to the disparity between the flamelet time scales and the residence times of formation of the slow forming species, restricting the application of the multiple unsteady flamelet modeling. Solving the unsteady flamelet allows preserving the history of change in scalar dissipation over time and space and hence can model the non-equilibrium effects. Since the unsteady diffusion flamelet model (UDFM) solves the chemistry in one-dimensional form, it is considerably faster compared to other models like Eddy Dissipation Concept (EDC), laminar finite rate or PDF Transport models which treat kinetics in same dimensions as flow.

The present formulation is applied in the similar fashion as explained in the previous work of the [9, 12] and the UDFM is assumed to have little impact on average temperature and mass fraction of global species, as compared to SDFM. This allows using UDFM as a post-processing activity with frozen flow field. The UDFM solves unsteady marker probability Eq. (2) on a steady converged solution using SDFM. The transport equation for the unsteady flamelet marker probability $I$ is:

$$\frac{d}{dt}(\rho I_n) + \nabla \cdot (\rho \vec{v} I_n) = \nabla \cdot \left[ \frac{\mu_t}{\sigma_t} \nabla I_n \right] \qquad (2)$$

Where, $I_n$ is the nth probability, which represents the average weight of particles corresponding to $n$ th level of scalar dissipation. The cumulative probability, which is the sum of all marker probabilities, is initialized as:

$$I = \sum_{n=1}^{N} I_n = \begin{bmatrix} 1 \rightarrow Z \geq Z_{init} \\ 0 \rightarrow Z < Z_{init} \end{bmatrix} \qquad (3)$$

Where $Z$ is the mean mixture fraction and $Z_{init}$ is the user supplied mixture fraction for initialization. The value of $I$ is set toward zero at the inlet boundaries and as the flamelet gets convected and diffused out of the domain the $I$ field decreases to zero with time. The initial flamelet field is calculated from a steady diffusion flamelet solution with all the slow forming species zeroed in the initial flamelet profile. In the case of multiple unsteady flamelets, the probability corresponding to nth flamelet is initialized inside a sub domain created using Eq. (4) as follows:

$$I_n = \begin{cases} 1 & if \quad \chi_{min,n} < \chi \leq \chi_{max,n} \\ 0 & if \quad \chi < \chi_{min,n}, \quad \chi > \chi_{max,n} \end{cases} \quad (4)$$

where $\chi_{min,n}$ and $\chi_{max,n}$ gives the scalar dissipation range for the nth unsteady flamelet. The scalar dissipation at stoichiometric mixture fraction is calculated as a probability weighted volume integral from the steady-state field at each time step as:

$$\chi_{st,n} = \frac{\int_V I_n \chi_{st}^{3/2} \rho dV}{\int_V I_n \chi_{st}^{1/2} \rho dV} \quad (5)$$

where V represents the fluid volume. The mean scalar dissipation at the stoichiometric mixture fraction is limited to a maximum value approximately equal to the extinction scalar dissipation of the flamelet. The flamelet temperature $T^{ad}(Z,t)$ is calculated at the probability weighted scalar dissipation $\chi_{st}$ at each time step from the steady diffusion flamelet library.

The mean species mass fractions as predicted by the unsteady flamelet over time are given as:

$$\tilde{Y}_k = \frac{\sum_n \iint_{f,t} I_n(t) \, Y_k(f,t) \, p(f) df \, dt}{\sum_n \int_t I_n(t) \, dt} \quad (6)$$

Where, $Y_k(Z,t)$ denotes the mass fraction of the $k$ th species and $p(Z)$ represents the $\beta-\text{PDF}$. The flamelet equations for the species mass fractions are solved until the probability marker has convected and diffused out of the domain substantially.

### 1.2.1 Eddy Dissipation Concept approach

In the second approach, the turbulence-chemistry interaction is modeled using Eddy Dissipation Concept model [24], which was developed from the Eddy Dissipation Model of [25] to take into the account of chemical kinetics. It is based on the energy cascade model and gives an expression for mean reaction rate. The EDC model assumes that molecular mixing and chemical reactions occur in the fine structures of the order of Kolmogorov length scales in the turbulent flow field corresponding to the smaller dissipative eddies. Their characteristic length $L^*$ and velocity $u^*$ is defined as:

$$L^* = \frac{2}{3}\left(\frac{3C_{D2}^3}{C_{D1}^2}\right)^{1/4}\left(\frac{\upsilon^3}{\tilde{\varepsilon}}\right)^{1/4} \quad (7)$$

$$u^* = \left(\frac{C_{D2}}{3C_{D1}^2}\right)^{1/4}(\upsilon\tilde{\varepsilon})^{1/4} \quad (8)$$

where $\nu$ is the kinematic viscosity and $C_{D1}$=0.134 and $C_{D2}$=0.5 [24]. The mass-averaged filtered reaction rate for the nth species is given as follows:

$$-\bar{\omega}_n = \frac{\bar{\rho}\dot{m}\chi}{1-\gamma^*\chi}(\tilde{Y}_n - Y_n^*) \quad (9)$$

where $\chi$ is the reacting fraction of the fine structures and is considered as 1 in the present study as suggested by [26]. $\tilde{Y}_n$, the mass fraction of each computational cell is solved using the species mass transport equation for each individual species. For the calculation of the $Y_n^*$, corresponds to the fine structures, is calculated using a system of ODE's representing a Perfectly Stirred Reactor (PSR) system:

$$\frac{dY_n^*}{dt} = \frac{w_n^*}{r^*} + \frac{1}{t^*}(Y_n^0 - Y_n^*) \tag{10}$$

$w_n^*$ is calculated from the chemical kinetic mechanism. Further, PSRs are assumed to be at the steady state, which is achieved by integrating the Eq. (10) in time till the steady state is achieved.

**Description of Test Cases**

**1.3 Sandia Flames**

Sandia flames are a series of six piloted methane-air jet flames (Flame A to F) stabilized on Sydney burner [27-28] and have been used as a part of the TNF workshop (http://www.sandia.gov/TNF/). The chosen test cases for this study are the flames series D-F. These flames have been studied in detail by [27, 29]. [27] made simultaneous measurements of major and minor species, temperature, and mixture fraction whereas [29] performed detailed velocity measurements. Since these flames have increasing velocity in their fuel and pilot jets, they have an increasing probability of local extinction in the downstream section of the flame length. Flame D has a very small probability of local extinction whereas extinction becomes more prominent in flames E and F due to higher jet and pilot velocities ratios.

The main jet of the burner has an inner diameter of 7.2 mm and is surrounded by a wall of 0.25mm thickness. The fuel jet is surrounded by an annular region of the pilot with an outer diameter of 18.2mm which consists of 72 tiny premixed jets that produce a homogeneous temperature and gas distribution [29]. A wall of 0.35mm surrounds the pilot annulus which separates the pilot region from the laminar annular co-flow of air at 1.0m/s. The main jet burns a mixture of three parts of air and one part of $CH_4$ by volume. This mixture is used to reduce the problem of fluorescence interference due to soot, and also to reduce the flame length and produce a much more robust flame than pure $CH_4$ [28]. The pilot is a lean ($\varphi = 0.77$) mixture of $C_2H_2$, $H_2$, air, $CO_2$, and $N_2$ with the same nominal enthalpy and equilibrium composition as methane/air at this equivalence ratio [27]. The stoichiometric mixture fraction of the flame based on a mixture fraction of unity in the fuel jet is 0.351. The flow rates to the main jet and pilot are scaled in such a way that the energy release of the pilot is approximately 6% of the main jet for each flame [30].

## 2. Numerical Details

In the present work, ANSYS FLUENT 16.0 (Ansys 2015) and OpenFOAM 2.1 [32] are used to perform the simulations using the unsteady flamelet modeling and eddy dissipation concept approach, respectively, for all the flames.

In ANSYS FLUENT, the pressure-velocity coupling scheme used is PISO and least square cell-based approach is used to calculate the gradients required for constructing the values of scalars at cell faces and computing secondary diffusion terms and velocity derivatives. The pressure interpolation scheme used is PRESTO and second order upwind and second order implicit schemes are used for the convective fluxes and transient formulation of the unsteady flamelet respectively.

In the case of OpenFOAM, rhoPisoFOAM [8] solved based on the compressible URANS modeling, is used. The linear-upwind-interpolation scheme is applied to calculate all the convective fluxes whereas second-order central differences approximation is used for the diffusion fluxes. The dynamically adjustable time stepping in conjunction with second-order implicit Euler method is used for time integration to maintain courant number less than the value of 0.4. The integration of the mean reaction rate in each computational cell is solved using the RADAU5 algorithm, designed for solving the stiff chemistry by implementing 5th-order accurate implicit Runge-Kutta method and RADAU quadrature formula. The relative tolerance, absolute tolerance and a maximum number of iterations are set to $5 \times 10^{-5}$, $1 \times 10^{-5}$ and 107, respectively, in order to achieve the desired accuracy. Turbulence is modeled using a standard k-epsilon model with default turbulent constants. The steady laminar flamelet model (LFM) is used to model the turbulent-chemistry interaction to obtain a converged steady flow solution and then the Eulerian unsteady laminar flamelet model is used to model the slow kinetic species as a post processing on the steady-state, steady flamelet solution. The detailed description of the approach can be found in [13, 4].

## 2.1 Modeling Approach

The burners considered for the present study have concentric annular geometry and the flames developed are generally axisymmetric in nature. Hence, 2D axisymmetric grids are used for the present computational work. The grid developed is orthogonal and non-uniform with moderate expansion in the axial as well as a radial direction having finer resolution towards the jet and pilot nozzle. A computational domain of dimension 110 $D_j$ x 30 $D_j$ is used for the Sandia series flames, shown in fig. 1. Grid independence study of the flames is performed using two grids, one coarser and the other finer. The coarser grid has 350 X 130 cells in the axial and radial directions whereas the finer mesh has 500 X 150 cells in the case of Sandia flames. Since the predictions of the grids are in good agreement, the coarser grid is used for the rest of the simulations for all Sandia flames. The results of the grid independence study for all the flames are shown in fig. 2.

## 2.2 Boundary Conditions

For the Sandia flames used in the present study, the scalar data are available at the TNF workshop archives and the complete 2-component velocity data are available as a part of the TNF workshop. The flow parameters from the experimental database at the inlet boundary are summarized in Table 1. The experimental data available for these flames [28] include profiles at an axial distance of 1mm from the jet exit and so the mean axial and radial profiles of velocity at the inlet are explicitly specified using this data. This method has been used earlier in the works of [18]. Turbulence is specified using turbulence intensity and hydraulic diameter. A free stream turbulence intensity of 1% is used for the calculations [33]. Reportedly, there is negligible difference between the burnt gas composition of the pilot mixture and that of the $CH_4$/air mixture at the same total enthalpy and equivalence ratio. Hence, the pilot composition is taken as that of an unstrained $CH_4$/air flame ($\varphi = 0.77$) with a stoichiometric mixture fraction of 0.27. The temperature of the jet is 290 K and that of the pilot is 1880 K for the D and E flames. The temperature of the pilot is slightly lower for the flame F [33] and hence a temperature of 1860 K is used for the present calculations as suggested by [34].

3. **Results and Discussion**

In this section, the results obtained from the unsteady diffusion flamelet modeling (UDFM) and eddy dissipation concept (EDC) approach are presented in parallel for individual Sandia flame and are further compared with the experimental measurements to explain the deviations. The comparison of the velocity, turbulent kinetic energy, mixture fraction, temperature, and species mass fraction predictions, using fairly detailed GRI mechanism is performed to study the effect of the chemical kinetics and turbulence-chemistry interactions. In the last section, the investigation of the dominance of the different NO formation pathways, majorly, prompt, thermal, nitric oxide and NNH routes, have been discussed.

### 3.1 Flow Field and Flame Structure

#### 4.1.1 Velocity and turbulent kinetic energy predictions

The radial distributions of the axial velocity for all the Sandia flames series, i.e. D to F, chosen for the present study are shown in Fig. 3. The steady flamelet model has been proven to predict the velocity, temperature and mixture fraction fields with good agreement with the experimental data by various previous studies on piloted flames [1, 21]. Since the unsteady flamelet equation is solved over a converged steady flamelet solution as a post-processing step with a frozen flowfield, a comparison of the steady and unsteady flamelet predictions of the flow field is not required. An acronym UDFM and EDC are used from here onwards to represent unsteady diffusion flamelet modeling (UDFM) and eddy dissipation concept (EDC) approaches, respectively. A good description of the velocity flow field is a prerequisite for the accurate predictions of reactive scalars as well as flame length and position.

Simulated results at three different axial locations at $x/D$ = 15, 30 and 45 from the fuel jet exit, are compared for all the three Sandia flames. In the vicinity of the fuel jet nozzle i.e. $x/D$ = 15, the magnitude of the centerline axial velocity is accurately captured by the finite rate based approach (EDC), whereas in the case of UDFM, an under-prediction of approximately 3.6 %, 6.1%, and 6.3% are observed in D, E and F flames, respectively, and is depicted in Fig. 3. However, at the downstream region of the higher reaction thickness zone, i.e. at $x/D$ = 30, the centerline axial velocity magnitude of the flame D using UDFM exceeded EDC by a merely 3%. Whereas, under-prediction of approximately 6% and 15% observed with EDC and UDFM, respectively, in the case of flame F, which can be attributed to the increased ratios of the jet as well as pilot velocities. The experimental velocity database for $x/D$ = 30 is not available in the case of the flame E. Interestingly at $x/D$ = 45, prediction close to the centerline from EDC approach are performing better as compared to UDFM approach for flame D and E, where an under-prediction of approximately 30% is observed with UDFM. For flame F, the under-prediction is consistent at 20% for both EDC and UDFM approaches. However, distributions in the radial extent of all the flames are more scattered i.e. higher radial spreading rate, in the vicinity of the fuel jet and both the approaches are able to reasonably capture the further downstream progression of the axial velocity. The radial over-spreading of the jet is likely due to the over-prediction of the turbulent kinetic energy in the proximity of the fuel jet. However, the predictions using both the approaches are improving while traversing in the downstream direction for the flame D. Whereas, over-prediction in the turbulent kinetic energy is quite prominent in the downstream regions of the flame F and can be easily seen in the Fig. 4.

**4.1.2 Mixture fraction predictions**

Figure 5 displays the radial distributions of the mixture fraction at three different axial locations for all the flames. In the vicinity of the fuel jet i.e. at $x/D = 15$, mixture fraction is analogously predicted, with a marginal difference for all the flames but over-spreading in the radial extent is observed. However, at $x/D = 30$, an extensive decrement of approximately 22% and 13% under EDC approach are observed along the centerline for flame D and E, whereas, accurate prediction of the magnitude is captured for flame F, indicating the slow progression of the mixture fraction field under high degree of local extinction. On the other hand, for UDFM approach, the maximum deviation of approximately 5% and 9% are observed for the flame D and E, and an increase of 5% in magnitude is observed along the centerline axis for flame F. Although the deviation of mixture fraction predictions near the centerline at $x/D = 30$ is less when compared to the results obtained by [18], the over predictions in the outer regions of the radial profiles suggest an early radial overspreading of the jet. The mixture fraction prediction of flame

D in the present study are more accurate than the CMC predictions observed by [16] and are comparable to the results obtained by [12] using EPFM although much more agreeable results using LES were observed by [35]. Further, UDFM approach is accurately predicting the mixture fraction distribution at x/D = 45 in the case of flame F, whereas consistent under-prediction is observed in the mixture fraction field under EDC approach.

### 4.1.3 Temperature predictions

The temperature field plays a vital role in capturing the NO formation zones due to its strong dependence on the formation rates, which thereby restricting it to a very narrow region in the vicinity of the maximum temperature [36]. The Favre averaged temperature predictions are compared with the experimental measurements in Fig. 6. The temperature field along the centerline and close to the fuel jet, i.e. at x/D = 15, is well captured for flame D using UDFM and peak value is over-predicted by 150K. Whereas with the EDC approach, the peak of the maximum temperature in the reaction shear layer over-predicted by 500K. This peak corresponds to the similar location in the case of flame E and F, the over-prediction extended to approximately 240K and 580K for UDFM approach and 750K and 1050K for EDC approach. However, the radial extent of the temperature distributions is well captured using both the turbulence-chemistry interaction approaches. At the further downstream location, i.e. x/D = 30, over-prediction of the temperature magnitude of approximately 250K, 410K, and 560K is observed along the centerline of the axisymmetric computational domain with UDFM approach in flame D, E, and F respectively. Furthermore, with EDC approach, this over-prediction augmented to 450K, 520K, and 700K, respectively for flame D, E, and F. Surprisingly at x/D = 45, the temperature field with EDC approach is in good match with the experimental measurements for all the three flames whereas under-prediction of nearly 150K is observed along the centerline of the flame D and an over-prediction of 320K and 280K are observed in the case of flame E and F. The thickened reaction shear layer in case of EDC approach clearly illustrates the reason for the over-prediction in the temperature magnitude in the radial distributions and can be seen from the contours shown in Fig. 7.

The mean temperature profile near to the centerline is slightly over predicted and temperature decrease with increase in strain rate is observed as can be expected in the case of adiabatic flames. This global over prediction can be expected as the adiabatic flamelet modeling neglects radiative heat losses. This adiabatic assumption is seen more prominent in the EDC approach shown in Fig. 7, where flame length is increased for the flame E and F. The peak value of mean temperature at x/D = 15 is over predicted in flame F the most with the predictions being close to the experimental data in the case of flame D. This is consistent with the fact that highest probability of extinction is reported at x/D = 15 and 30 for flame F. The results agree well with the experimental data near the centerline at downstream locations (x/D = 45) for all the flames suggesting healthy reignition with a swift recovery of the flame. Flame extinction and the reignition could influence the turbulence fields by affecting the heat release and in turn affect the velocity and mixture fraction fields.

### 4.1.4 Intermediate and minor species profiles

The axial profiles of CO, $H_2$ and OH mass fractions for all the flames are compared with the experimental database as these species are most influenced by the transient effects [12] apart from NO which is studied in detail computationally, in the later sections. The discrepancies in the predictions of the species mass fractions can be attributed to the ones observed in the scalar and velocity predictions. The centerline distribution of the intermediate CO mass fraction is shown in Fig. 8. The intermediate CO is over-predicted in the rich part of the flame using EDC for all the three cases. This can be accredited to the poor mixing time scale associated with the EDC approach as the evolution of CO corresponds to the frozen-like chemistry [37]. The predictions of CO mass fraction with UDFM are in reasonable agreement with the experimental database for all the three flames.

The axial distribution of the $H_2$ species is also presented in Fig. 8. The predictions are well captured for flame D and E using UDFM, whereas, significant over-prediction in the maximum value and broad spectrum of the distributions is seen more pronounced for flame F. This can be attributed to the increase in the probability of flame extinction from flame D to F and also to the increase in over prediction of mean temperatures in this region. Nevertheless, the results with EDC approach are highly over-predicted as compared to the experimental measurements as well as predictions from the UDFM approach. The strong non-linearity of the species evolution in the combustion systems leads to challenges in the quantification of the OH radicals [38]. Also, OH radicals can be regarded as the reaction rate indicator. The maximum peak by UDFM approach over-predicted by a factor of approximately 1.5, 1.3, and 1.53 in the flame D to F, respectively. However, under-prediction of exactly 1.5 is observed in the case of flame D using EDC approach, whereas, significant increase in OH mass fraction is seen in the case of the flame E and F. This can be attributed to the influence of stronger turbulent mixing with increased level of mixing from simple diffusion flame D to partially premixed with strong extinction and re-ignition dependent flame E and F. The current predictions of intermediate and minor species mass fractions of flame D with GRI mechanism are comparable to EPFM predictions by [12] although the CMC predictions on the same flame by [16] managed to capture the peak value of species mass fractions slightly closer to the experimental data with similar chemistry mechanism.

## 3.2 NO predictions

NOx refers collectively to all the oxides of nitrogen that are formed during combustion at high temperatures of which the most hazardous are NO and $NO_2$. However, NO is the principal species in NOx formation and species like $NO_2$ and other oxides of nitrogen together contribute to less than 5 percent of the total NOx. The experimental database available for the flames includes only NO mass fraction designating as NOx emission. As a result, the present study is focused on predicting NO mass fraction from the three piloted flames and other oxides of nitrogen are addressed during the post-processing in order to diagnose the most prominent pathway for NO formation.

Fundamentally, the major NO formation in a gaseous combustion system can be due to three mechanisms namely thermal NOx, prompt NOx, and NOx via $N_2O$ intermediates. Prompt NOx usually forms in regions of high-speed reactions occurring at the flame front due to the reaction of hydrocarbon radicals like C, CH, $CH_2$ etc. with atmospheric nitrogen. The Fenimore mechanism representing prompt NOx formation is significant in the fuel rich regions of the flame and involves the intermediate formation of HCN and further its oxidation to NO [39].

$$CH + N_2 \leftrightarrow HCN + N$$
$$C + N_2 \leftrightarrow CN + N \tag{15}$$
$$N + OH \leftrightarrow NO + H$$

where the first reaction is considered as the most significant contributor of prompt NOx. [40] performed flamelet/progress variable formulation of nitric oxide formation in Sandia D-flame and observed that major NO contribution originated from the prompt NOx mechanism due to the partially premixed fuel composition since dilution of the fuel resulted in lower flame temperatures and higher stoichiometric mixture fractions, reducing the efficiency of Zeldovich mechanism. Alternatively, NO can form via $N_2O$ intermediates and is important in the fuel lean low-temperature conditions although the contribution due to this is considerably less compared to the Zeldovich mechanism. A detailed description regarding the different routes of formation and reduction of NOx and the reactions involved in them can be found in [40]. The predicted results of the NOx in the Sandia flames in conjunction with GRI mechanism are discussed in the following session. The assessment of the two turbulence-chemistry interaction approaches in calculating NOx formation will serve as one of the primary objective of the present study as traversing from varying from simple diffusion flame (flame D) to the partially premixed flame with a higher degree of extinction (flame E and F).

### 3.2.1 Predictions of NO mass fraction

The axial and radial distributions of the NO mass fraction along the centerline and at various axial locations, respectively, are compared with the experimental data are shown in Figs. 9 and 10. The current chemistry mechanism (GRI 3.0) includes the chemistry for NO formation via thermal, prompt and nitrous oxide mechanisms, since the study by [16] on Sandia Flame D reported that a skeletal mechanism including only thermal NOx formation chemistry under predicted the NO mass fraction significantly and concluded that prompt NOx mechanism is one of the dominant routes for formation of NO in the methane-air diffusion flames. In the case of UDFM approach, the maximum value of NO mass fraction is highly over-predicted as compared to the experimental measurements by approximately a factor of 2, 2.5 and 3.4 in the flame D, E, and F, respectively, and is shown in Fig. 9. The NO mass fraction of nearly double the experimental measurements is also observed in the axial extent of the domain, most probably due to the higher residence time scales associated with the NO formation in this fuel rich region. Further, the early formation of NO species is seen in the case of the flame E and F, which can be ascribed to over-predicted temperature magnitude in the corresponding region, shown in Fig. 9. Overprediction of NO has been reported by various previous modeling studies including unsteady flamelet modeling [6, 12] and CMC [16] modeling, between locations x/D = 15 to 45 in Sandia flame D which are consistent with the results observed. This prediction is seen highly augmented in the case of EDC approach by a substantial factor of 3, 18 and 17 in the flame D, E, and F, respectively. Similar to UDFM, the over-prediction with further higher discrepancies are seen in the axial extent of the domain with EDC approach. Although the radiant fraction is relatively small in the Sandia flames D-F, [23] observed that the radiative transfer due to gas-molecular radiation becomes important at downstream locations at x/D > 40. [16] observed the highest NO increase for the adiabatic modeling of flame D using GRI mechanism and [40] also observed that considering the radiation in the optically thin limit resulted in the reduction of temperature by 70K and a reduction in NOx by 20% at x/D = 45. Thus, the overprediction at larger downstream distances can be attributed to the negligence of gas molecular absorption and emission of radiation to an extent. NOx formation mechanism involves complex chemistry with a large number of elementary reactions describing the formation and reburn of oxides of nitrogen.

For a comprehensive view of the early NO formation using both the turbulence-chemistry interaction approaches, the radial distributions at four different axial locations are shown for all the flames in the Fig. 10. Due to thickened shear layer between fuel rich and lean side of the flame in the vicinity of the fuel jet, i.e. at x/D = 3, maximum NO formation is quite prominent in the case of EDC approach. However, UDFM approach is able to predict the reasonable maximum mass fraction as compared to the experimental measurements. Nevertheless, the location of the maximum NO mass fraction is accurately captured in all the flames. The maximum NO mass fraction remains nearly consistent while moving from the flame D to E, whereas reduction of approximately 20% in the NO mass fraction is seen in the shear layer in the flame F, which can be attributed to the higher momentum issued by the pilot and fuel jet. Progressing in the downstream direction at x/D = 15, the predictions are highly over-predicted in the case of EDC approach, by approximately a factor of 7, 8.5, and 40 in the flame D, E and F, respectively. Later in further downstream location, this over-prediction in maximum NO mass fraction using EDC approach is approximated at a factor of 4 and 15 for flame D and E, respectively, whereas an exorbitant increase by a factor of 24 is seen in the case of flame F. However, relatively, a reasonable prediction is achieved with UDFM approach i.e. predictions are approximated at a factor of 1.75, 2.5, and 4.5 for flame D, E, and F, respectively, depicting the advantage of UDFM over EDC approach in capturing the slowly evolving NO species. Further, In the region close to maximum temperature, i.e. x/D = 45, the maximum value of the NO mass fraction is approximately over-predicted by a factor of 2.25, 4.4, and 26, for the EDC approach and a factor of approximately 1.7, 2.6, and 3, for the UDFM approach, arranged in the order of flame D, E, and F, respectively. These predictions are evident from the contour distribution of the NO mass fraction shown in the Fig. 11, where the thickened shear layer with high NO content can be easily distinguished.

### 3.2.2 **NOx Controlling Pathways**

The major concern which should be addressed now is the formation of different NO containing species in the maximum NO mass fraction zone, i.e. essentially in the reaction layer in the context of EDC and UDFM approach. To underline this study, a maximum mass fraction of the different pathways is tracked during the post processing in order to quantify the dominance of the reaction pathways in the NO formation. Previously, [41] have quantified the dominance of different NO formation pathways in Sandia flame D and observed that nearly half of the total NOx produced is formed by means of a prompt pathway. Whereas, rest of the NOx is distributed by $N_2O$ and thermal pathway, in the descending order. The present study is concentrated on differentiating the impact of prompt, thermal, nitric oxide and NNH route for participating in the NOx formation, especially, in the environment of increasing local extinction in the flames. Fuel NOx is not considered in the present study. Figure 12 signifies the maximum mass fraction of HCN radicals captured in the flame brush. HCN radicals are primarily responsible for NOx formation in the high-temperature flame brush through a prompt mechanism [41] and intermediate species NCO is also considered which is formed under reaction of the HCN radicals with free radicals' species such as O and OH. The enormous difference can be seen between the maximum HCN mass fraction predicted by EDC and UDFM approach. Predictions from EDC approach are approximately 110%, 290%, and 275% higher as compared to the UDFM approach, for the flame D, E, and F, respectively. The over-prediction of HCN radicals is in agreement with the highly over-predicted NO formation using EDC approach. However, the maximum mass fraction of NCO radicals' species predicted from both EDC and UDFM approach is consistently three orders lesser as compared to the HCN radicals'. The alteration in the maximum mass fraction of the NCO is less than 5% for all the flames, which indicates the subsidiary significance of NCO radicals in the NOx formation via the prompt mechanism.

Usually, the conventional method [46] is adopted in the calculation of the thermal NO formation. The thermal NO is obtained by subtracting the net NO calculated in each flame from the prompt NO considered in the present study. Thermal NO is highly dependent on the accurate predictions of the temperature field across the flame brush and due to over-predicted temperature field with EDC approach may lead to higher production of the thermal NO mass fraction. Figure 13 shows the non-linear sensitivity of the maximum thermal NO mass fraction obtained from the conventional method under EDC and UDFM approach. For flame D, the maximum thermal NO mass fraction obtained from the EDC approach shows discrepancies of around 50% as compared to the UDFM. These discrepancies with EDC approach intensified by nearly 700% and 550% for the flame E and F. However, one order magnitude difference is observed in the maximum thermal NO mass fraction as compared to the HCN radical's mass fraction in the prompt mechanism with the EDC approach. Comparatively, the increment with the UDFM approach is barely much amplified. [40] pointed out NO formation via $N_2O$ mechanism is considerably much smaller (approximately 5%) as compared to that of the Zeldovich mechanism. Nitric oxides are mostly formed under fuel-lean and high-pressure conditions. $N_2O$ is formed significantly in the shear layer between fuel and oxidizer, resulting into consumption zone in the core of the flame. For flame E and F, EDC approach presents increment of one order of magnitude as compared to the UDFM approach, whereas, the moderate rise is seen in the case of flame D, shown in Fig. 14. The NNH route is defined by [42] primarily resulting from the two mechanisms:

$$N_2 + H \Rightarrow NNH$$

$$NNH + O \Rightarrow NO + NH$$

Similar to the $N_2O$ flame environment dependency, [43] suggested that NOx formation through NNH pathway are predominantly governed in the lean premixed and low-temperature combustion systems. This can be demonstrated by the range of the maximum NNH and NH mass fraction, which is at least four and three order lesser than the HCN and thermal pathways, respectively. This clearly signifies that thermal NO pathway shows higher sensitivity towards the NOx formation, followed by the prompt, nitric oxide and NNH pathways. However, the predictions from the EDC approach are highly over-predicting which can be easily depicted from the flame structure and flow field evolution.

Whereas the UDFM approach is able to predict the NO formation with the reasonable accuracy, and the results are in good agreement with the experimental measurements.

## 4. CONCLUSIONS

In this work, the numerical investigation has been carried out using UDFM for piloted turbulent diffusion flames to study NOx formation using detailed chemistry. The effect of two different turbulence-chemistry interaction approaches are used to investigate the NOx formation in the turbulent jet piloted flames varying from simple diffusion flame (flame D) to the partially premixed flame (Flame E and F) with a higher degree of local extinction. The chemical kinetics is studied using the fairly detailed mechanism with NOx chemistry namely GRI 3.0. The flow field defining quantity, i.e. velocity, turbulent kinetic energy are in reasonable agreement in the downstream regions with the experimental measurements, whereas elongated flame structure is observed with EDC as compared to the UDFM approach. The intermediate species CO and $H_2$ are well predicted by the UDFM approach whereas over-prediction is observed in the case of OH species radial distribution. The NO predictions increases from flame D to E and F, however, the increment of the over-prediction factor in the case of flame E and F are not that pronounced among themselves. The slowly developing NO mass fraction is well predicted by the UDFM approach, however, exorbitant over-prediction is observed with the EDC approach, depicting the capability of the UDFM approach in calculating slowly developing NO chemistry. Among the different pathways for the NO formation in the flame D, E, and F, the dominance of these different pathways is seems to be performing in the descending order leading by thermal NOx, Prompt NOx, $N_2O$ route, and NNH route. Similar to the substantial over-prediction of the NO mass fraction from the EDC approach, the immense increase in the HCN mass fraction (Prompt NO) and thermal NO mass fraction is seen with the EDC approach. The effect of radiative heat losses with gray and non-gray radiation in the NO levels and accuracy of radiation models in predicting NO mass fractions using NOx models in these flames is a potential field of further study.

ACKNOWLEDGMENT

The authors would like to acknowledge the IITK computer center (www.iitk.ac.in/cc) for providing the resources to perform the computation work, data analysis, and article preparation.

| Flame | $Re_{jet}$ | $U_{jet}$ - cold | $U_{pilot}$ - burnt | $U_{coflow}$ |
|---|---|---|---|---|
| D | ~22,400 | 49.6 | 11.4 | 0.9 |
| E | ~ 33,400 | 74.4 | 17.1 | 0.9 |
| F | ~44,800 | 99.2 | 22.8 | 0.9 |

Table 1: Experimental flow parameters of Sandia flames series used in the present study.

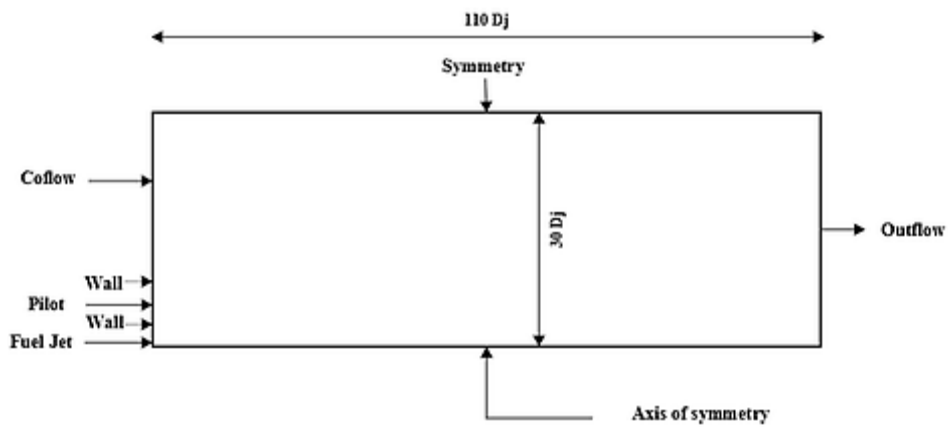

Fig 1: Schematic and computational domain with imposed boundary conditions for Sandia flames.

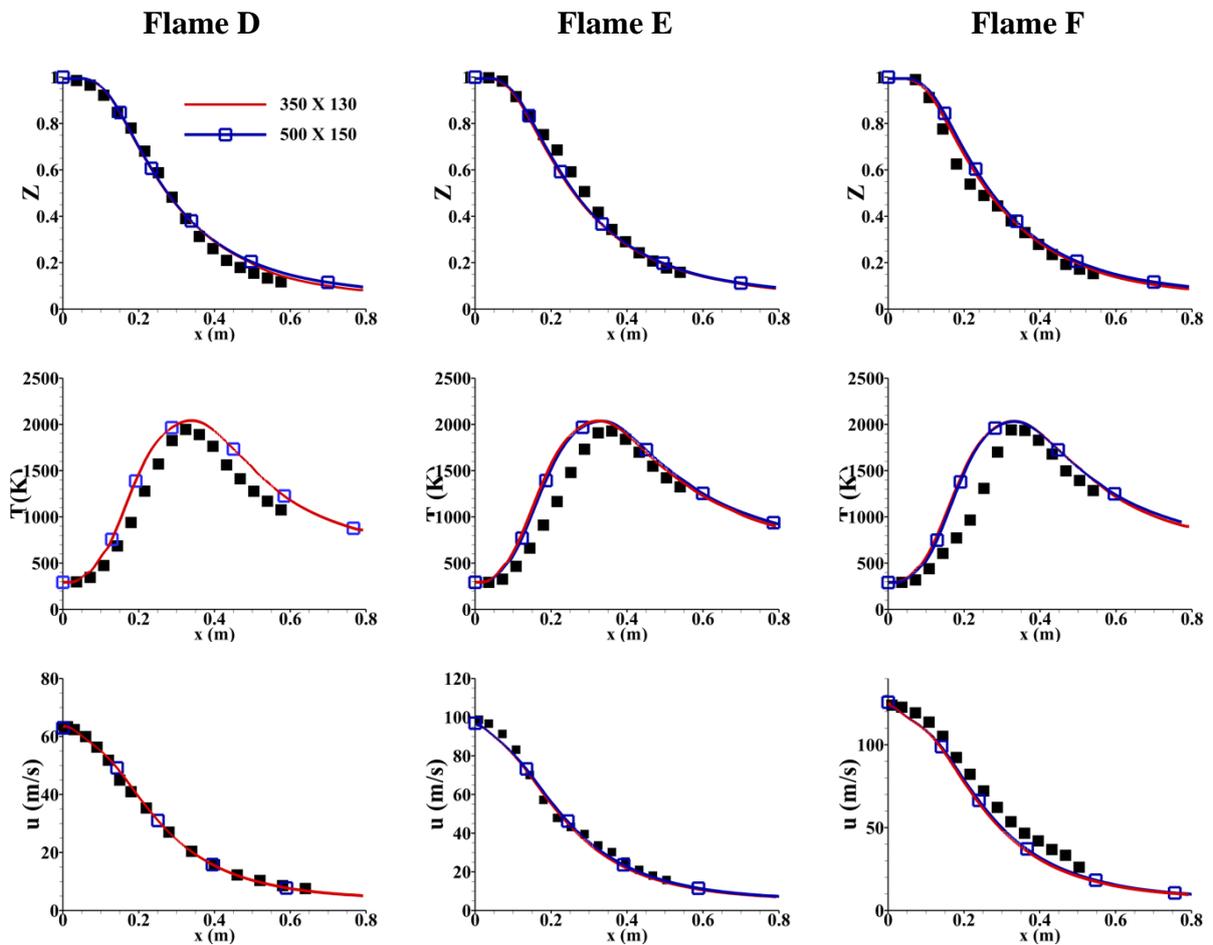

Figure 2: Grid independence study showing the axial distributions of the mixture fraction (Z), temperature (T) and axial velocity (u) for two different non-uniform grids in case of the flame D, E, and F, respectively, shown in column wise fashion.

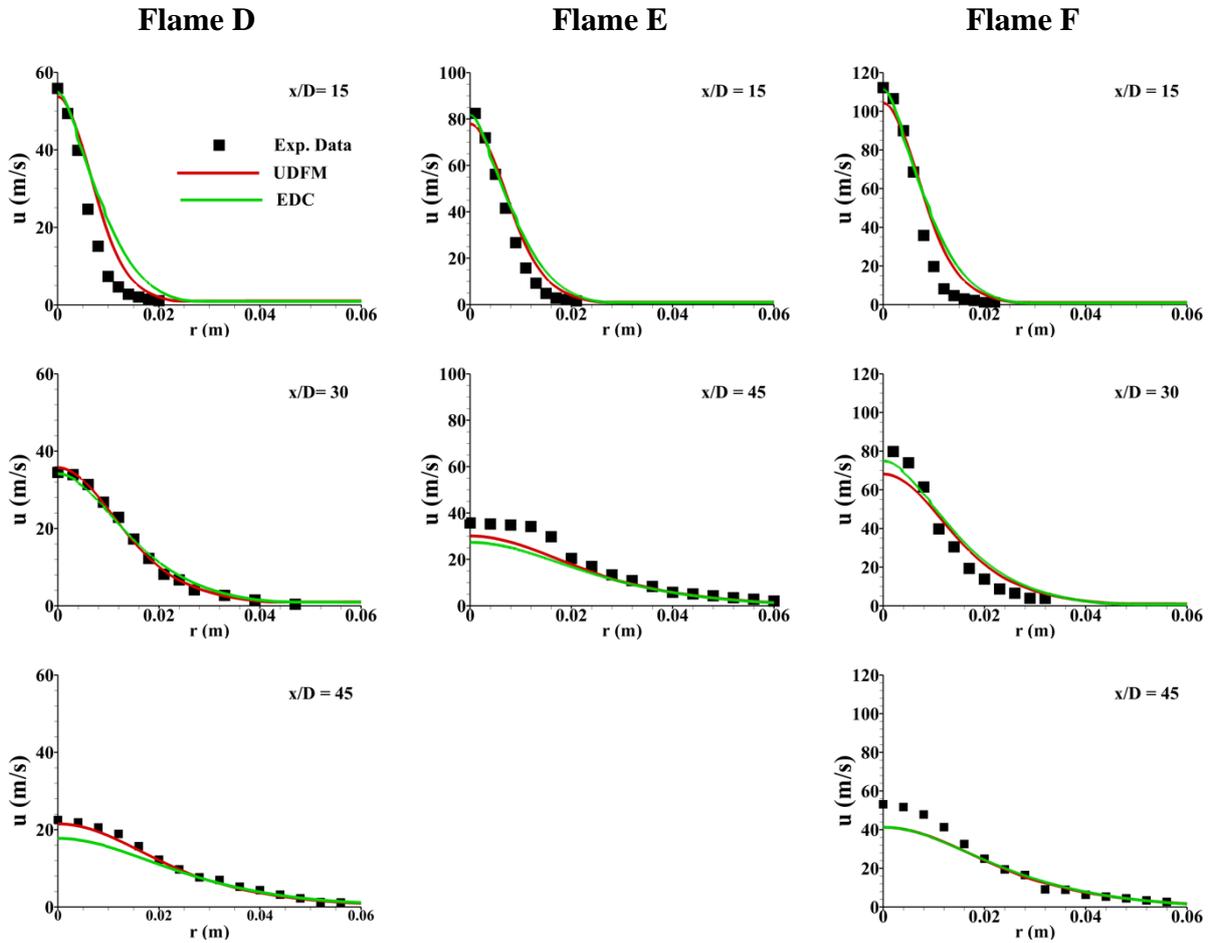

Figure 3: Radial distributions of the axial velocity (u) at three different axial locations from the fuel jet inlet for Sandia flames (D, E, and F) using two different turbulence-chemistry interaction approaches.

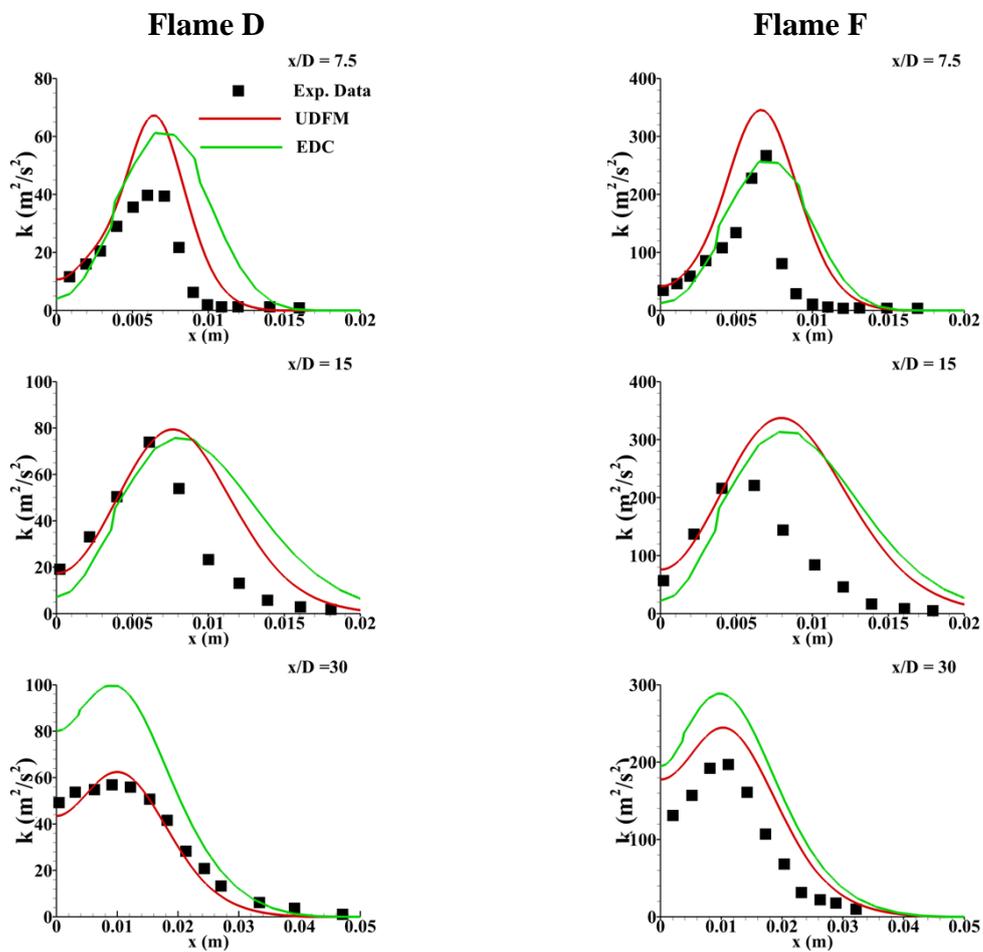

Figure 4: Effect of two different turbulence chemistry interaction approaches on the radial distribution of the turbulent kinetic energy (k) for Sandia flame D and F

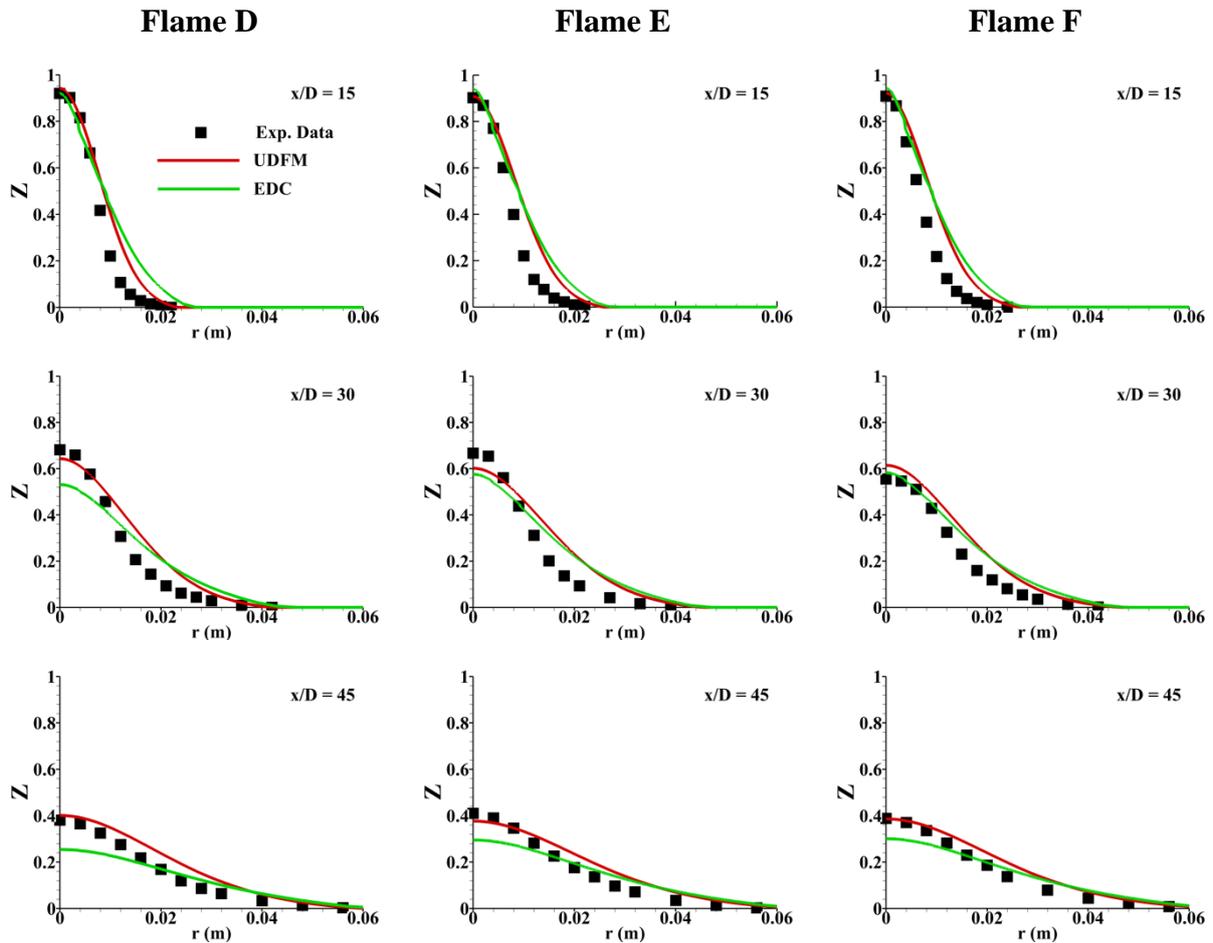

Figure 5: Radial distribution of the mixture fraction (Z) at three different axial locations for Sandia flames (D, E, and F) with two different turbulence-chemistry interaction approaches.

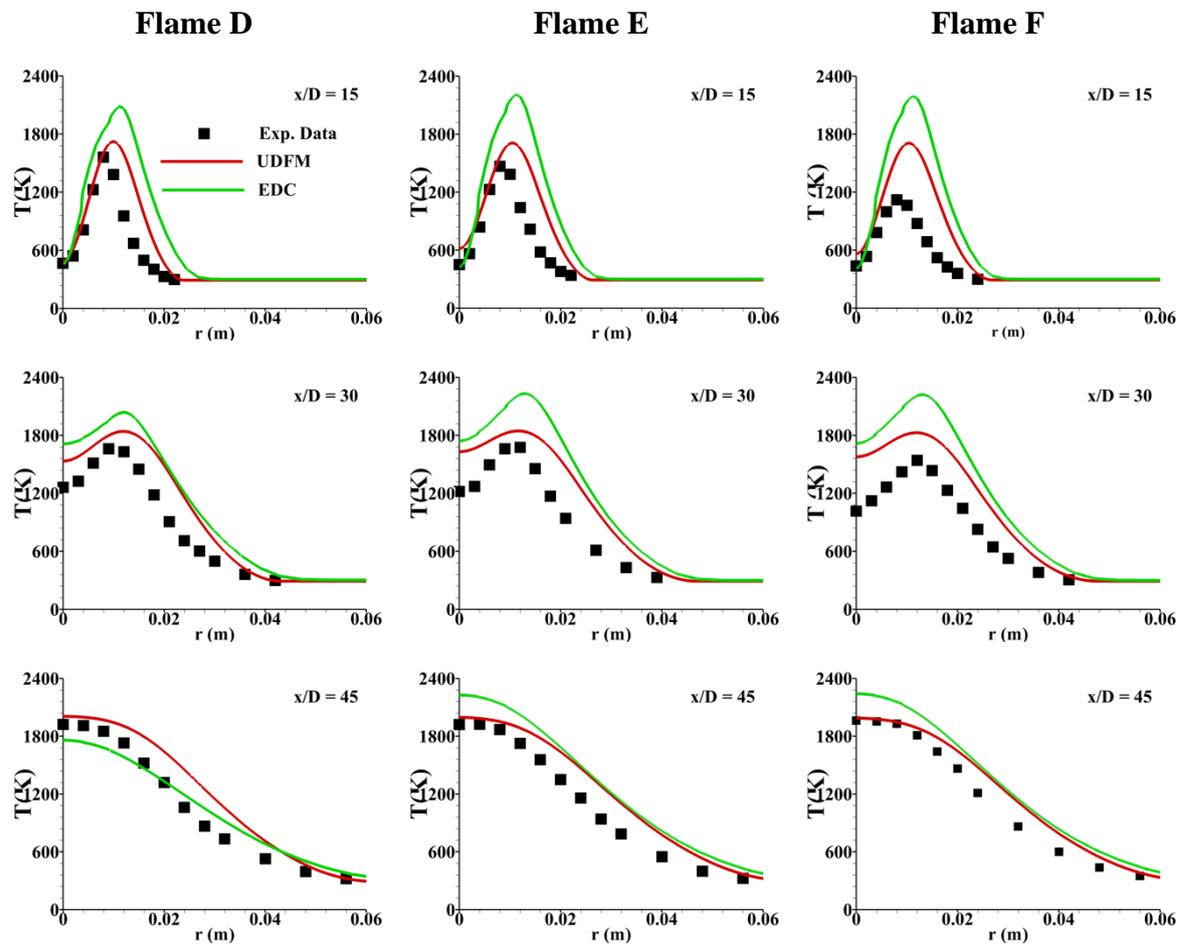

Figure 6: Impact of UDFM and EDC approach on the radial evolution of the temperature field at three different axial locations for the Sandia flame D, E and F.

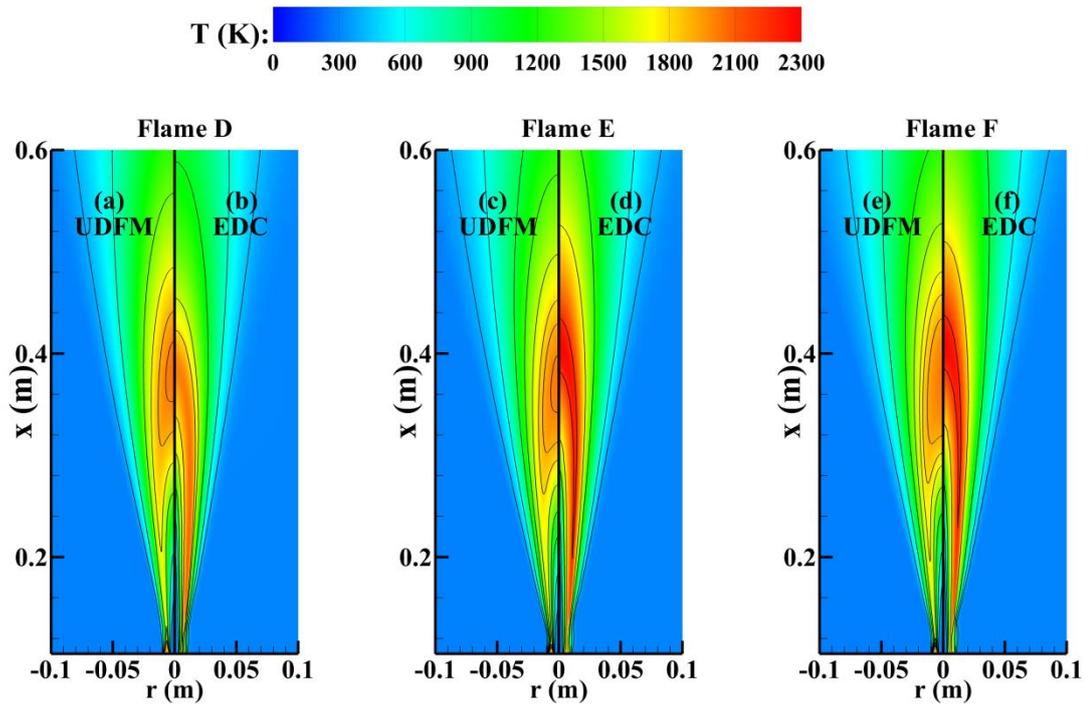

Figure 7: Temperature contours showing the effect of the UDFM and EDC approach on the spatial evolution of the Sandia flame D, E, and F.

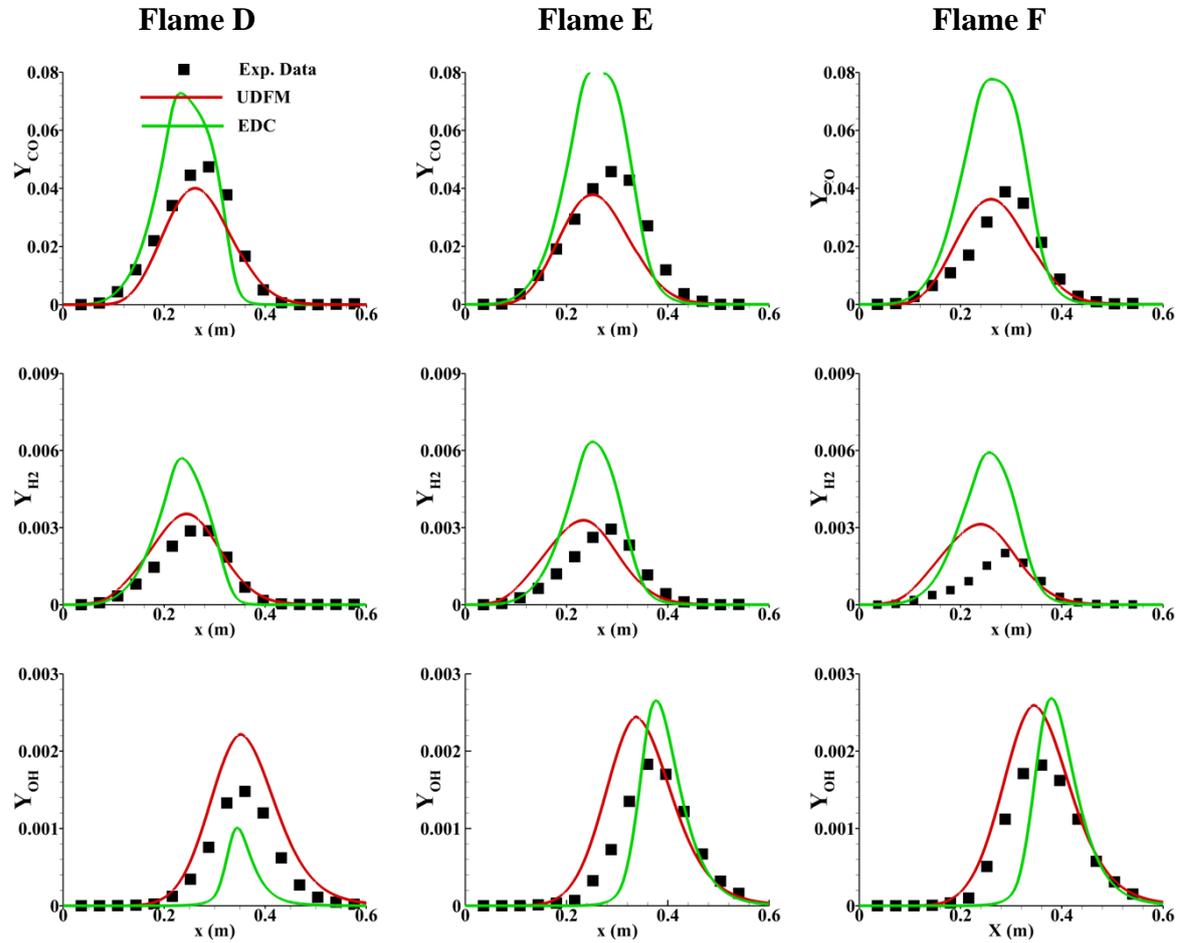

Figure 8: Axial distributions along the centerline axis of the species mass fractions of CO (1st column), H2 (2nd column) and OH (3rd column) for Flame D, E, and F.

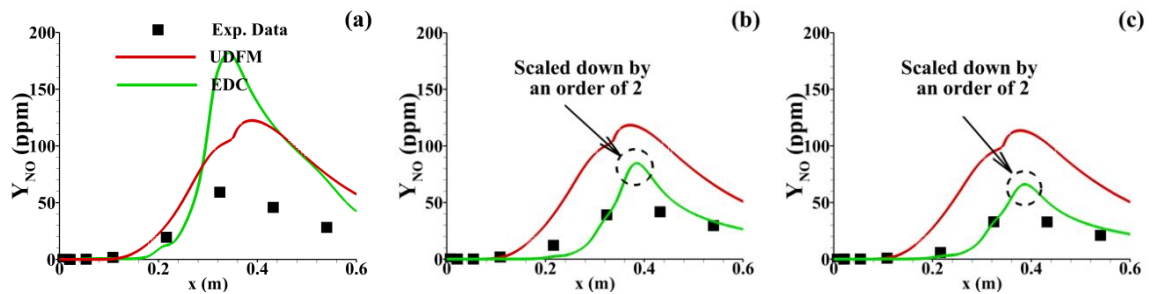

Figure 9: Axial distribution along the centerline axis of the NO mass fraction for Sandia (a) flame D, (b) flame E and (c) flame F; Symbols are experimental measurements.

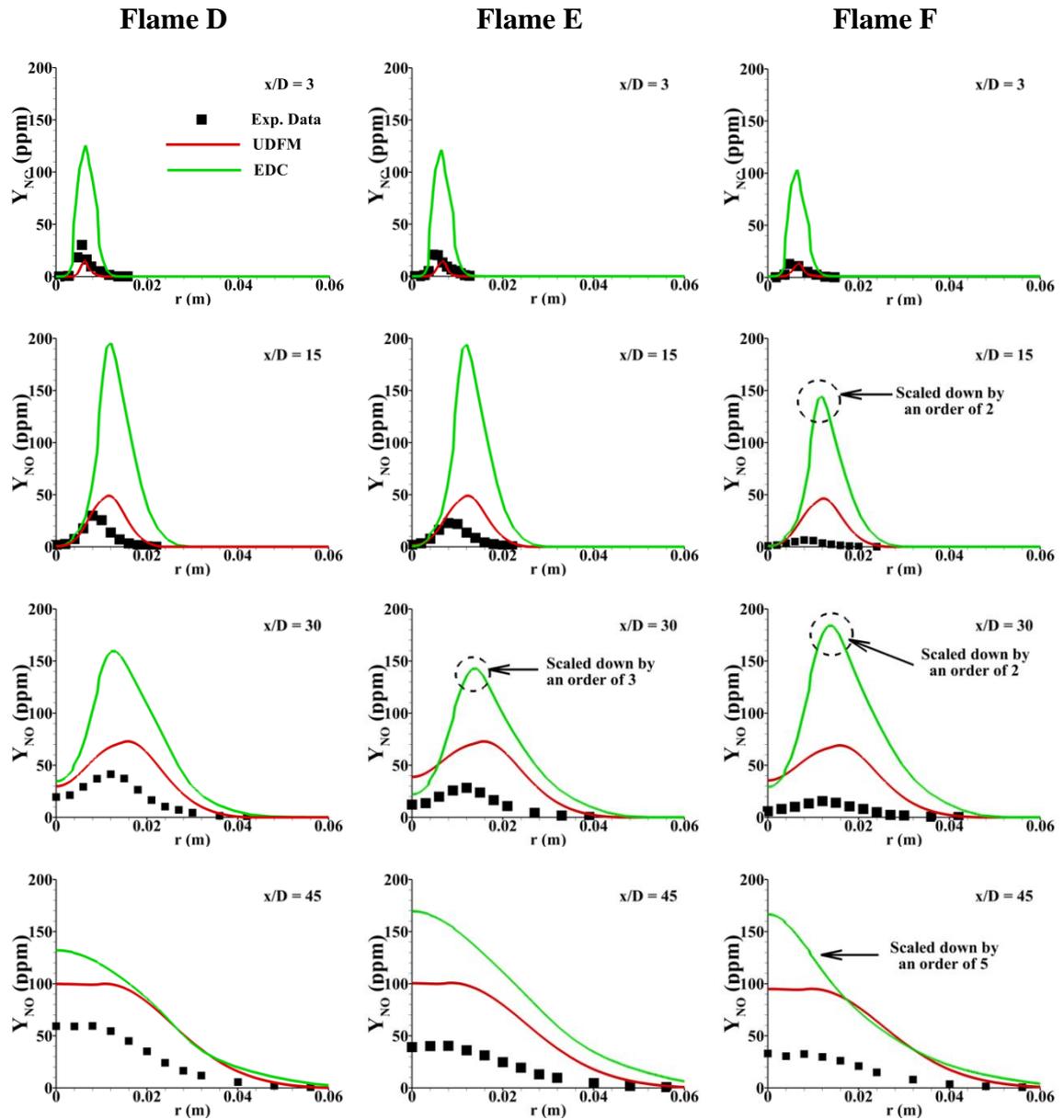

Figure 10: Influence of the UDFM and EDC approach on the NO mass fraction at three different axial location of the Sandia flame D, E, and F.

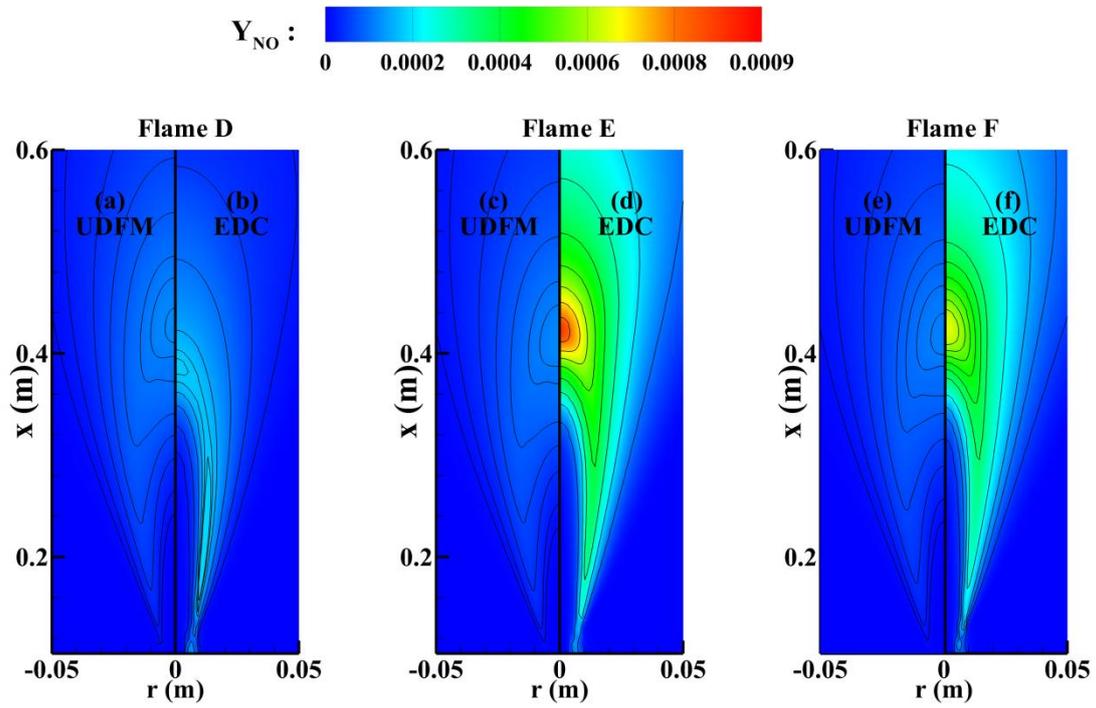

Figure 11: Contours of the NO mass fraction showing the effect of the UDFM and EDC approach for Sandia flame D, E, and F.

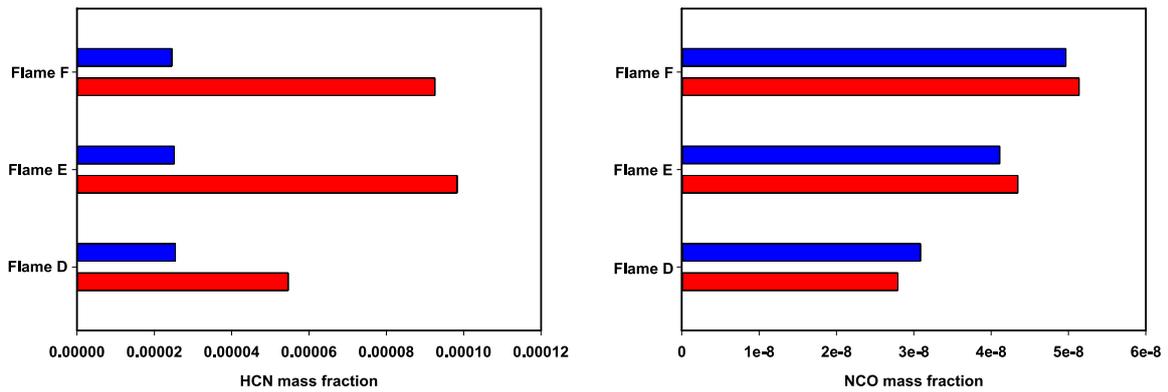

Figure 12: The bar-graph distribution of the maximum HCN and NCO mass fraction in the flame brush, accounting the NO formation through Prompt mechanism, in the flame D, E, and F. The blue and red bars correspond to the UDFM and EDC approach, respectively. This bar color convention remains the same in all subsequent plots (Figs. 13-15) unless or otherwise mentioned.

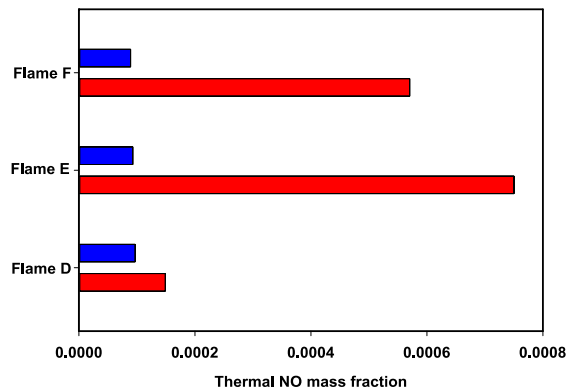

Figure 13: The bar-graph distribution of the maximum Thermal NO mass fraction produced during the combustion process of the flame D, E, and F.

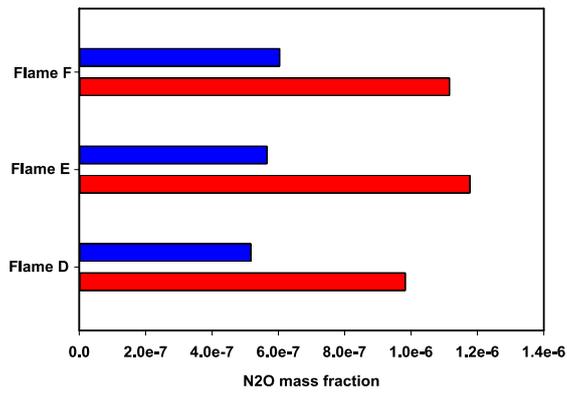

Figure 14: The bar-graph distribution showing the variation of the maximum $N_2O$ mass fraction produced through nitric oxides route of the NO formation.

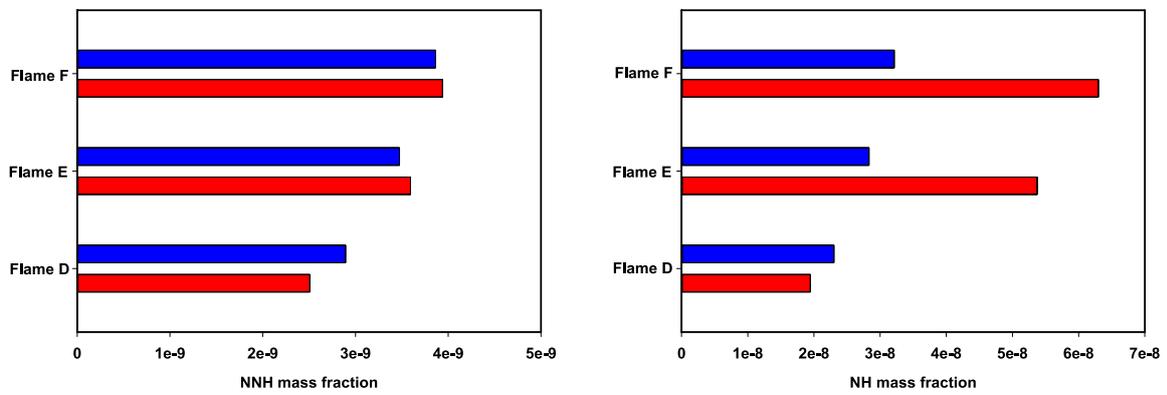

Figure 15: Evolution of the maximum NNH and NH mass fraction through NNH route of the NO formation in the flame D, E, and F.